\documentclass[aps,prd,eqsecnum, showpacs]{revtex4}

\usepackage{epsf}

\usepackage{psfrag}

\begin{document}
\author{M. Giammatteo}
\email{mgiammatteo@hotmail.com}
\affiliation{School of Mathematics and Statistics,\\University of Newcastle upon Tyne,\\Newcastle upon Tyne NE1 7RU U.K.}
\title{Dirac quasinormal frequencies in Schwarzschild-AdS space-time}
\author{Jiliang Jing}
\email{jljing@hunnu.edu.cn}
\affiliation{Institute of Physics and Department of Physics,\\Hunan Normal University, Changsha, Hunan 410081 P. R. China}

\date{ November 2004}

\begin{abstract}
We investigate the quasinormal mode frequencies for the massless
Dirac field in static four dimensional $AdS$ space-time. The
separation of the Dirac equation is achieved for the first time
in $AdS$ space. Besides the relevance that this calculation can
have in the framework of the $AdS/CFT$ correspondence between
M-theory on $AdS_4\times S^7$ and SU(N) super Yang-Mills theory
on $M_3$, it also serves to fill in a gap in the literature, which
has only been concerned with particles of integral spin $0,1,2$.
\end{abstract}
\maketitle

\section{Introduction}
Perturbations of black holes are dominated, over intermediate
timescales, by characteristic modes known as quasinormal modes.
They are similar to normal modes of a closed system, but since
the field can fall into the black hole or radiate to infinity,
the modes decay and the corresponding frequencies are complex
\cite{chandra}. They have been extensively studied in
asymptotically flat spacetimes and good reviews on the topic can
be found in \cite{nollert}, \cite{kokkotas}. The study of quasinormal modes in Anti-de Sitter space was first done in \cite{mann}.

More recently quasinormal frequencies have been investigated
in the context of string theory in Anti-de Sitter space
\cite{horowitz}, \cite{lemos}, \cite{moss}. There is a
suggestion, known as $AdS/CFT$ correspondence, that
string theory in Anti-de Sitter space is equivalent
to a conformal field theory in one dimension fewer
\cite{maldacena}. In the framework of this conjecture
the study of $AdS$ black holes has a direct interpretation
in terms of the dual conformal field theory on its boundary.
 The duality predicts that the retarded $CFT$ correlation
 functions are in one to one correspondence with Green's
 functions on Anti-de Sitter space with appropriate boundary
 conditions \cite{danielsson}, \cite{kalyana}, \cite{birmingham},
 \cite{son}. Furthermore, as mentioned in \cite{horowitz}, it
 is assumed that a large static black hole in $AdS$ space
corresponds to a thermal state in the $CFT$ on the boundary.
Perturbing the black hole is equivalent to perturbing this
thermal state. The perturbed system is expected, at late times,
to approach equilibrium exponentially with a characteristic
time-scale. This time-scale is inversely proportional to the
imaginary part of the poles of the correlators of the
perturbation operator. It seems that these relaxation time-scales
are quite complicated to calculate in the $CFT$, therefore their
computation is conveniently replaced by the evaluation of the
quasinormal frequencies in the $AdS$ bulk space. In \cite{lemos}
and \cite{moss}, the authors went beyond the scalar perturbation
treated in \cite{horowitz} and they considered electromagnetic
and gravitational perturbations. In one of the most recent works
on the subject, Berti and Kokkotas \cite{berti} confirm and
extend previous results on scalar, electromagnetic and
gravitational perturbations of static $AdS$ black holes and
analyse, for the first time, Reissner-Nordstrom Anti-de Sitter
black holes and calculate their quasinormal frequencies.

The interaction of a Dirac field with a black hole has been studied
by Finster and his collaborators in a series of papers. While they
have found stable particle-like solutions in the Einstein-Dirac-
Maxwell system \cite{finster1, finster2}, they have also proved
the non-existence of time-periodic solutions in various black hole
space-times \cite{finster4, finster5}. This means that Dirac
particles, including electrons and neutrinos, cannot remain on a
periodic orbit around a black hole. In \cite{finster3}, they also
showed that there are no spherical symmetric black hole solutions
in the Einstein-Dirac-Maxwell system other than the
Reissner-Nordstr{\"o}m one. This suggests that if a cloud of Dirac
particles undergoes gravitational collapse, the fermionic particles
either vanish inside the event horizon of a black hole or escape to
infinity.

The quasinormal frequencies related to the evolution of a massless
Dirac field in a Schwarzschild black hole space-time, were studied
in \cite{cho}. The quasinormal modes of the Reissner-Nordstr{\"o}m
de Sitter black hole for Dirac fields were studied by using the
P{\"o}shl-Teller potential approximation in Ref. \cite{Jing}. In
this paper we are interested in the analysis of the modes of
vibration of a massless Dirac field in a Schwarzschild Anti-de
Sitter background space and it is for this reason that we compute
the related quasinormal mode frequencies. This calculation is
relevant to the $AdS/CFT$ correspondence between M-theory on
$AdS_{4}\times S^{7}$ and $SU(N)$ super Yang-Mills theory on
$M_3$, and  it also serves to fill in a gap in the literature,
which, in $AdS$ spaces, has only been concerned with particles of
integral spin $0, 1, 2.$

\section{Dirac equation in static AdS space-time}
\label{section2}
The metric for a Schwarzschild Anti-de Sitter black hole can be
written as
\begin{eqnarray}
ds^2=-f dt^2+\frac{1}{f}dr^2+r^2(d\theta^2 +sin^2\theta d\varphi^2),
\end{eqnarray}
with
\begin{equation}
f=1-\frac{2M}{r}-\frac{\Lambda}{3}r^2,
\end{equation}
where the parameters $M$, and $\Lambda$ represent the black hole mass,
 and the negative cosmological constant, respectively.

The Dirac equation in a general background space-time can be written,
according to \cite{Brill}, as
\begin{equation}\label{Di}
[\gamma^a e_a{}^\mu(\partial _\mu+\Gamma_\mu)]\Psi=0.
\end{equation}
Here, $\gamma^a$ are the Dirac matrices,
\begin{equation}
\gamma^{0}= \left(
\begin{array}{cc}
-i&0\\0&i
\end{array}\right),\ \ \
\gamma^{i}=\left(
\begin{array}{cc}
0&-i\sigma^{i}\\i\sigma^{i}&0
\end{array}\right),\ i=1,2,3,
\end{equation}
while $\sigma^{i}$ are the Pauli matrices,
\begin{equation}
\sigma^{1}= \left(
\begin{array}{cc}
0&1\\1&0
\end{array}\right),\ \ \
\sigma^{2}=\left(
\begin{array}{cc}
0&-i\\i&0
\end{array}\right),\ \ \
\sigma^{3}=\left(
\begin{array}{cc}
1&0\\0&-1
\end{array}\right).
\end{equation}
The four-vectors $e_a{}^\mu$ represent the inverse of the tetrad
$e_a{}^\mu$ defined by the metric $g_{\mu\nu}$ as,
\begin{equation}
g_{\mu\nu}=\eta_{ab}e^{a}{}_{\mu}e^b{}_{\nu},
\end{equation}
with $\eta_{ab}={\rm diag}(-1, 1, 1, 1)$ being the Minkowski metric.
$\Gamma_\mu $ are the spin connection coefficients, which are given
by
\begin{eqnarray}
\Gamma_\mu= \frac{1}{8}[\gamma^a,\gamma^b]e_a{}^\nu e_{b\nu;\mu}.
\end{eqnarray}
Here, $e_{b\nu;\mu}=\partial_{\mu}e_{b\nu}-\Gamma^{\alpha}_{\mu\nu}
e_{b\alpha}$ is the covariant derivative of $e_{b\nu}$ and $\Gamma^
{\alpha}_{\mu\nu}$ is the Christoffel symbol.

We take the tetrad to be
\begin{eqnarray}
e_\mu^a={\rm diag}(\sqrt{f}, \frac{1}{\sqrt{f}}, r, r \sin \theta).
\end{eqnarray}
The spin connection $\Gamma_\mu $ can then be expressed as
\begin{eqnarray}
\Gamma_0&=&\frac{1}{4}f' \gamma_0\gamma_1, \nonumber \\
\Gamma_1&=&0, \nonumber \\
\Gamma_2&=&\frac{1}{2}\sqrt{f}\gamma_1\gamma_2, \nonumber \\
\Gamma_3&=&\frac{1}{2}(\sin\theta
\sqrt{f}\gamma_1\gamma_3+\cos\theta \gamma_2\gamma_3 ).
\end{eqnarray}
The Dirac equations (\ref{Di}) become
\begin{eqnarray}
-\frac{\gamma_0}{\sqrt{f}}\frac{\partial \Psi}{\partial t}+\sqrt{f}
\gamma_1 \left(\frac{\partial }{\partial r}+\frac{1}{r}+\frac{1}{4 f}
\frac{d f}{d r} \right) \Psi+\frac{\gamma_2}{r}(\frac{\partial }
{\partial \theta}+\frac{1}{2}\cot\theta)\Psi+\frac{\gamma_3}{r
\sin\theta}\frac{\partial \Psi}{\partial \varphi}=0. \label{Di1}
\nonumber\\
\end{eqnarray}
If we re-scale $\Psi$ as
\begin{eqnarray}
    \Psi=f^{-\frac{1}{4}}\Phi,
\end{eqnarray}
Eq. (\ref{Di1}) assumes a simpler form in the new unknown $\Phi$,
which can be written as
\begin{eqnarray}
    -\frac{\gamma_0}{\sqrt{f}}\frac{\partial \Phi}{\partial t}+
    \sqrt{f} \gamma_1 \left(\frac{\partial }{\partial r}+\frac{1}{r}
    \right) \Phi+\frac{\gamma_2}{r}(\frac{\partial }{\partial \theta}
    +\frac{1}{2}\cot\theta)\Phi+\frac{\gamma_3}{r \sin\theta}
    \frac{\partial \Phi}{\partial \varphi}=0. \label{Di2}
\end{eqnarray}
We introduce a well known coordinate change from the radial variable
$r$ to the tortoise coordinate $r_{*}$ given by
\begin{eqnarray}
    r_*=\int \frac{d r}{f}.
\end{eqnarray}
We will use an ansatz for the Dirac spinor
\begin{eqnarray}\label{C5an}
    \Phi=\left(
\begin{array}{c}
\frac{i G^{(\pm)}(r)}{r}\phi^{\pm}_{jm}(\theta, \varphi) \\
\frac{F^{(\pm)}(r)}{r}\phi^{\mp}_{jm}(\theta, \varphi)
\end{array}\right)e^{-i \omega t},
\end{eqnarray}
with spinor angular harmonics
\begin{eqnarray}\label{C54}
    \phi^{+}_{jm}=\left(
\begin{array}{c}
\sqrt{\frac{j+m}{2 j}}Y^{m-1/2}_l \\
\sqrt{\frac{j-m}{2 j}}Y^{m+1/2}_l
\end{array}\right), \ \ \ \ \ \ \ \ \ \ \ \  (for \ \
j=l+\frac{1}{2}),
\end{eqnarray}
\begin{eqnarray}
    \phi^{-}_{jm}=\left(
\begin{array}{c}
\sqrt{\frac{j+1-m}{2 j+2}}Y^{m-1/2}_l \\
-\sqrt{\frac{j+1+m}{2 j+2}}Y^{m+1/2}_l
\end{array}\right), \ \ \ \ \ \ (for \ \ j=l-\frac{1}{2}).
\end{eqnarray}
Note that $Y_{l}^{m\pm 1/2}(\theta,\varphi)$ represent ordinary
spherical harmonics. Since
\begin{eqnarray}
\left(\begin{array}{cc} -i\left(\frac{\partial}{\partial
\theta}+\frac{1}{2}\cot\theta \right) & \frac{1}{ \sin\theta}
\frac{\partial}{\partial \varphi}
\\ - \frac{1}{ \sin\theta}
\frac{\partial}{\partial \varphi} &
i\left(\frac{\partial}{\partial \theta}+\frac{1}{2}\cot\theta
\right)
\end{array}\right) \left(
\begin{array}{c}\phi_{jm}^{\pm} \\ \phi_{jm}^{\mp}\end{array}\right)=
i\left(\begin{array}{cc}
k_{\pm} & 0 \\ 0 & k_{\pm} \end{array}\right) \left(
\begin{array}{c}\phi_{jm}^{\pm} \\ \phi_{jm}^{\mp}\end{array}\right),
\end{eqnarray}
equations \ref{Di2} can be written in the simplified matrix form
\begin{eqnarray}
\left(\begin{array}{cc}0  & -\omega \\ \omega  & 0
\end{array}\right) \left(
\begin{array}{c}F^{\pm} \\G^{\pm}\end{array}\right)-\frac{\partial}
{\partial
r_*}\left(\begin{array}{c}F^{\pm}
\\G^{\pm}\end{array}\right)+\sqrt{f}\left(\begin{array}{cc}
\frac{k_{\pm}}{r}  & 0 \\ 0
&  -\frac{k_{\pm}}{r} \end{array}\right) \left(
\begin{array}{c}F^{\pm} \\G^{\pm}\end{array}\right)=0.
\end{eqnarray}
These equations can be thought of as equations referring to the
radial functions $(F^{+}, G^{+})$ or $(F^{-}, G^{-})$ depending on
the original choice made in the ansatz of equation \ref{C5an}. On
the other hand it can be proved that the radial differential
equations related to the two different sets of functions are
exactly the same. In other words, we have a unique equation for
$F^{\pm}$ and another equation for $G^{\pm}$. For this reason we
will avoid from now on specifying if we are referring to the spin
up or spin down components when talking about $F^{\pm}$ and
$G^{\pm}$ and we will simply call these functions $F$ and $G$.
Consequently the two decoupled equations can be expressed in the
form
\begin{eqnarray}
    \frac{d^2 F}{d r_*^2}+(\omega^2-V_1)F&=&0, \label{even}\\
    \frac{d^2 G}{d r_*^2}+(\omega^2-V_2)G&=&0, \label{odd}
\end{eqnarray}
with
\begin{eqnarray}
V_1&=&\frac{\sqrt{f}|k|}{r^2}\left(|k|\sqrt{f}+\frac{r}{2}\frac{d f}
{d r}-f\right), \ \ \ \left( k=j+\frac{1}{2},\ \ j=l+\frac{1}{2}
\right) \label{V1} \\
V_2&=&\frac{\sqrt{f}|k|}{r^2}\left(|k|\sqrt{f}-\frac{r}{2}\frac{d f}
{d r}+f\right), \ \ \ \left( k=-j-\frac{1}{2},\ \ j=l-\frac{1}{2}
\right). \label{V2}
\end{eqnarray}
In a similar way to the integer spin case in \cite{cho}, the two
potentials $V_1$ and $V_2$, are super-symmetric partners derived
from the same super-potential. In the asymptotically flat case,
it has been shown that potentials related in this way possess
the same spectra of quasinormal mode frequencies \cite{arlen}.
In the $AdS$ case, this is still true provided that the boundary
conditions are mapped between the two cases in a consistent way.
We shall concentrate just on Eq. \ref{even} with potential $V_1$
in evaluating the quasinormal mode frequencies in the next section.

\section{Dirac quasinormal frequencies}
\label{section3}
In this section we evaluate the quasinormal frequencies for the
massless Dirac field using the Frobenius series solution method.
We start defining $\Delta$ as
\begin{equation}
\Delta=r^2-2Mr+\alpha^2 r^4,
\end{equation}
where $\alpha^2 = -\Lambda/3$. This implies that the function $f$
in \ref{even} can be re-written as $f=\Delta/r^2$. Next, rescale
$F$ as
\begin{equation}\label{one}
F=e^{-i\omega r_{*}}u,
\end{equation}
which allows us to scale out the behaviour of the function at the
black hole event horizon. We obtain an equation in $u$ given by
\begin{equation}\label{three}
f^2 \frac{d^2 u}{dr^2}+\left\{f\frac{df}{dr}-2i\omega f\right\}
\frac{du}{dr}-Vu=0.
\end{equation}
Let us now set
\begin{equation}\label{four}
r=\frac{1}{x}.
\end{equation}
We obtain the equation in the variable $x$
\begin{equation}\label{SEVEN}
f^2 x^4 \frac{d^2 u}{dx^2}+fx^2 \left(2xf+x^2 \frac{df}{dx}+2i
\omega\right)\frac{du}{dx}-Vu=0.
\end{equation}
Define $p$ as
\begin{equation}
p=x^2-2Mx^3 +\alpha^2 x^4.
\end{equation}
Note that $p_1=p(x_1)=\Delta(x_1)x_{1}^4=0$. Rescale $f$ as
\begin{equation}\label{EIGHT}
f=\Delta x^2=px^{-2}
\end{equation}
inserting this into \ref{SEVEN} we obtain the equation
\begin{equation}\label{NINE}
p^2 \frac{d^2 u}{dx^2}+p\left(\frac{dp}{dx}+2i\omega\right)\frac{du}
{dx}-Vu=0.
\end{equation}
Rescale again the independent variable $x$, by introducing the
variable $z$ given by
\begin{equation}\label{TEN}
\left(\frac{x_{1}-x}{x_1}\right)=z^2,
\end{equation}
we obtain a new expression for $V$ in the variable $z$,
\begin{equation}
V(z)=p|k|^2+p^{1/2}|k|x_{1}\left(1-z^2\right)\left[3Mx_{1}
\left(1-z^2\right)-1\right].
\end{equation}
Insert \ref{TEN} into \ref{NINE} and rescale all the parameters: $M,
\alpha^2, \omega$ with respect to $x_1$, which is the inverse of the
radial coordinate $r_1$ of the black hole event horizon, while $r_1$
is the largest root of $\Delta$.
We obtain
\begin{equation}\label{tredici}
\hat{p}z\frac{d^2 u}{dz^2}-\left\{\hat p-\frac{d\hat p}{dz}z+
4i\hat\omega z^2\right\}\frac{du}{dz}-4z^3\left\{|k|^2+
\hat{p}^{-1/2}\left[3\hat M|k|\left(1-z^2\right)^2-
|k|\left(1-z^2\right)\right]\right\}u=0,
\end{equation}
where
\begin{equation}\label{quattordici}
\hat{p}=px_{1}^{-2}.
\end{equation}
Expanding the coefficients of equation \ref{tredici} as Taylor
series around $z=0$, which is the same than performing a series
expansion around the black hole event horizon, we obtain
\begin{equation}\label{quindici}
A(z)\frac{d^2 u}{dz^2}-B(z)\frac{du}{dz}-C(z)u=0,
\end{equation}
where
\begin{eqnarray}
A(z)&=&p_2 z^3 +p_3 z^5 +2\hat M z^7,\\
B(z)&=&(4i\hat\omega-p_2)z^2 -3p_3 z^4 -10\hat M z^6,\\
C(z)&=&4|k|^2 z^3 +2|k|p_2 z^2 \hat{q}^{-1/2} +4|k|p_3 z^4
\hat{q}^{-1/2}+12\hat M |k|z^6 \hat{q}^{-1/2},
\end{eqnarray}
with
\begin{equation}
p_2 =(6\hat M -2),
\end{equation}
\begin{equation}
p_3 =(-6\hat M+1)
\end{equation}
and
\begin{equation}
\hat{q} = \hat{p} z^{-2}=p_2+p_3 z^2 +2\hat M z^4.
\end{equation}
The differential equation in $u$ can be now solved by series by
exploiting Frobenius method, which consists in looking for a
solution of the form
\begin{equation}\label{C5S}
u=z^{\rho}\sum_{n=0}^{\infty}a_{n}z^{n}.
\end{equation}
The main difference between the spin $\frac{1}{2}$ case and the
integer spin case is the non-polynomial term $\hat{q}^{-1/2}$.
This function can be Taylor expanded as
\begin{equation}\label{sedici}
\hat{q}^{-1/2}=(p_2+p_{3}z^2 +2\hat{M}z^4)^{-1/2}=
\sum_{s=0}^{\infty}\frac{1}{s!}A_{s}z^{s},
\end{equation}
leaving aside for the moment the problem of deriving the
coefficients of this last series.

The two possible values for the index $\rho$ at $z=0$ of
our series solution \ref{C5S} can be easily calculated and
they are given by $\rho_{1}=0$ and $\rho_{2}=4i\hat\omega/p_2$.
We pick the first of the indices (the other solution has infinitely
many oscillations close to the horizon) and therefore look for a
series solution which will be simply given by
\begin{equation}\label{diciassette}
u=\sum_{n=0}^{\infty}a_{n}z^{n}
\end{equation}
The infinite nature of the expansion for $\hat{q}^{-1/2}$ and $u$,
leads us to deal with a product of two series which can be expressed
as a single series,
\begin{equation}\label{diciotto}
\left(\sum_{s=0}^{\infty}\frac{1}{s!}A_{s}z^{s}\right)
\left(\sum_{n=0}^{\infty}a_{n}z^{n}\right)=\sum_{n=0}^{\infty}
\left(\sum_{s=0}^{n}\frac{1}{s!}A_{s}a_{n-s}\right)z^{n}
\end{equation}
Inserting equations \ref{diciassette} and \ref{diciotto} into
equation \ref{quindici}, we obtain a recurrence relation for
$a_{n+1}$ which has $n$ terms,
\begin{eqnarray}
& &\left\{\left(n+1\right)\left[p_{2}\left(n+1\right)-4i\hat\omega
\right]\right\}a_{n+1}+\left\{p_{3}\left(n-1\right)\left(n+1\right)
-4|k|^2 \right\}a_{n-1}\nonumber\\
&+&\left\{2\hat{M}\left(n-3\right)\left(n+1\right)\right\}a_{n-3}
-2|k|p_{2}\left\{A_{0}a_{n}+\cdots+\frac{A_n}{n!}a_{0}\right\}\\
&-&4|k|p_{3}\left\{A_{0}a_{n-2}+\cdots+\frac{A_{n-2}}{(n-2)!}a_{0}
\right\}-12\hat{M}|k|\left\{A_{0}a_{n-4}+\cdots+\frac{A_{n-4}}
{(n-4)!}a_{0}\right\}=0\nonumber
\end{eqnarray}
This recurrence equation, due to its unbounded character caused by
the series expansion for $\hat{q}^{-1/2}$, is much harder to solve
by numerical recursion, and in any case before we proceed we first
need to know the expression for the coefficients $A_s$ of the Taylor
expansion of $\hat{q}^{-1/2}$. This problem is solved by remembering
Fa\`a di Bruno's formula for the analytical calculation of derivatives of any order
\begin{equation}\label{diciannove}
h_n=\left[\frac{d^{n}h}{dx^{n}}\right]_{x=x_0}=\sum_{k=1}^{n}f_{k}
\sum_{p(n,k)}n!\prod_{i=1}^{n}\frac{g_{i}^{a_{i}}}{(a_{i}\;!)(i\;!)
^{a_i}},
\end{equation}
where $h(x)=f[g(x)]$ is a composed function,
\begin{equation}
f_k=\frac{d^k}{dy^k}f(y_0)
\end{equation}
and
\begin{equation}
g_{i}=\frac{d^i}{dx^i}g(x_0).
\end{equation}
The second sum inside \ref{diciannove} is done over partitions
$p(n,k)$, which are defined as
\begin{equation}
p(n,k)=\left\{(a_1,\ldots,a_n):\;a_i\in\mathcal{N}_{0},\;
\sum_{i=1}^{n}a_{i}=k, \;\sum_{i=1}^{n}ia_i=n\right\}
\end{equation}
where $\mathcal{N}_{0}$ is equal to the set of non-negative
integers. An element $(a_1,\ldots,a_n)$ belonging to $p(n,k)$
represents a partition of a set with $n$ elements into $a_1$
classes of cardinality 1,$\ldots$, $a_n$ classes of cardinality $n$.
The number of such partitions is represented by the Stirling numbers
of the second kind,
\begin{equation}
S_{n}^{(k)}=\sum_{p(n,k)}n! \prod_{i=1}^{n}\frac{1}{(a_{i}\;!)(i\;!)
^{a_i}}.
\end{equation}
When formula \ref{diciannove} is applied to the $A_{s}$ coefficients,
we have
\begin{equation}
A_{s}=\sum_{m=1}^{s}f^{(m)}\left[\hat{q}(0)\right]\sum_{p(s,m)}s!
\prod_{i=1}^{s}\frac{1}{(a_{i}!)(i\;!)^{a_i}}\left\{\hat{q}^{(i)}
(0)\right\}^{a_i},
\end{equation}
with $f^{(m)}[\hat{q}(0)]$ given by
\begin{equation}
\prod_{k=0}^{m-1}\left(-\frac{(2k+1)}{2}\right)[\hat{q}(0)]^{-1/2-m}.
\end{equation}
On the other hand, evaluating $\hat{q}$ and its derivatives at $z=0$,
 we obtain
\begin{eqnarray}
\hat{q}(0)&=&6\hat{M}-2,\\
\hat{q}''(0)&=&2(1-6\hat{M}),\\
\hat{q}^{(4)}(0)&=&48\hat{M}
\end{eqnarray}
while
\begin{equation}
\hat{q}'(0)=\hat{q}'''(0)=0,
\end{equation}
beside all the derivatives of order higher than four. The $A_{s}$
coefficients can then be written in the form
\begin{equation}\label{C56}
A_s=\sum_{m=1}^{s}f^{(m)}\left[\hat{q}(0)\right]s!\left\{
\left[\frac{1}{(a_2 !)(2 !)^{a_2}}\left\{\hat{q}''(0)\right\}
^{a_2}\right]\left[\frac{1}{(a_4 !)(4 !)^{a_4}}\left\{\hat{q}
^{(4)}(0)\right\}^{a_4}\right]\right\},
\end{equation}
where the set of partitions $p(s,m)$ now contains only a single
element
\begin{equation}
p(s,m)=\left\{\left(a_2, a_4\right):\;a_2, a_4 \in
\mathcal{N}_{0},\;a_2+a_4=m,\;2a_2+4a_4=s\right\}.
\end{equation}
Hence
\begin{eqnarray}
a_{2}&=&2m-\frac{1}{2}s\\
a_{4}&=&\frac{1}{2} s-m,
\end{eqnarray}
for $4m>s>2m$ and $s$ even. Expression \ref{C56} is very simple to
encode in a numerical subroutine.

\subsection{Boundary conditions}
\label{BC}
Having separated Dirac equations in Anti-de Sitter Schwarzschild
black hole background spaces and obtained two radial second order
differential equations, \ref{even} and \ref{odd}, we have argued
that those two equations are related. They can be physically
identified as the equations governing the axial and polar
perturbations of the Dirac field, respectively. We have also
mentioned the fact that the two potentials \ref{V1} and \ref{V2} possess the same quasinormal mode frequency spectra if the boundary conditions are mapped consistently.

According to Cooper \cite{cooper}, equation \ref{even} describes
axial perturbations and equation \ref{odd} describes polar
perturbations of the Dirac field. Suppose we are interested
in investigating further the $AdS/CFT$ conjecture and  want
to relate the spinor field and its quasinormal frequencies
to the calculation of the correlation functions on the boundary
conformal field theory. More precisely, let us suppose that we
are interested in evaluating the correlators of a polar quantity
in the $CFT$ on the boundary, which corresponds to the polar part
of the spinor field in the bulk of $AdS$ space-time. Accordingly
we must set the {\it axial} spinor field perturbation to vanish
on the boundary.

In terms of what we have done until now, we must take our series
solution \ref{diciassette} to the axial equation \ref{even}, and
impose on this Dirichlet boundary conditions
\begin{equation}
u(z)=\sum_{n=0}^{\infty}a_{n}z^{n}\rightarrow 0
\end{equation}
as $z\rightarrow 1$. Taking the limit of the left hand side gives
us the identity
\begin{equation}\label{mainbc}
\sum_{n=0}^{\infty}a_{n}=0.
\end{equation}
This is an implicit equation for the computation of quasinormal
frequencies related to the spinor field. The computer program
which performs the numerical calculation is organized in the
following way. An independent routine evaluates the coefficients
$A_s$ associated to the Taylor expansion of $\hat{q}^{-1/2}$.
These results are then exploited within a recursive routine which
iterates the recurrence relation to evaluate the coefficients
$a_n$ of the series solution. Finally, an ordinary Newton-Raphson
root-finding routine seeks for solutions to the equation
\ref{mainbc}.

The choice of boundary conditions is related to the axial or polar
character of the spinor field at infinity and to its interpretation
in terms of the $AdS/CFT$ conjecture. On the other hand the
consistency of the boundary conditions can be checked by the shape
of the effective potential in the differential equation \ref{even}
we are solving.

\begin{figure}[ht!]
\begin{center}
\begin{tabular}{cc}
\psfrag{r}{$r$}
\psfrag{y}{$ $}
\epsfxsize=6.0cm
\epsfysize=6.0cm
\epsfbox{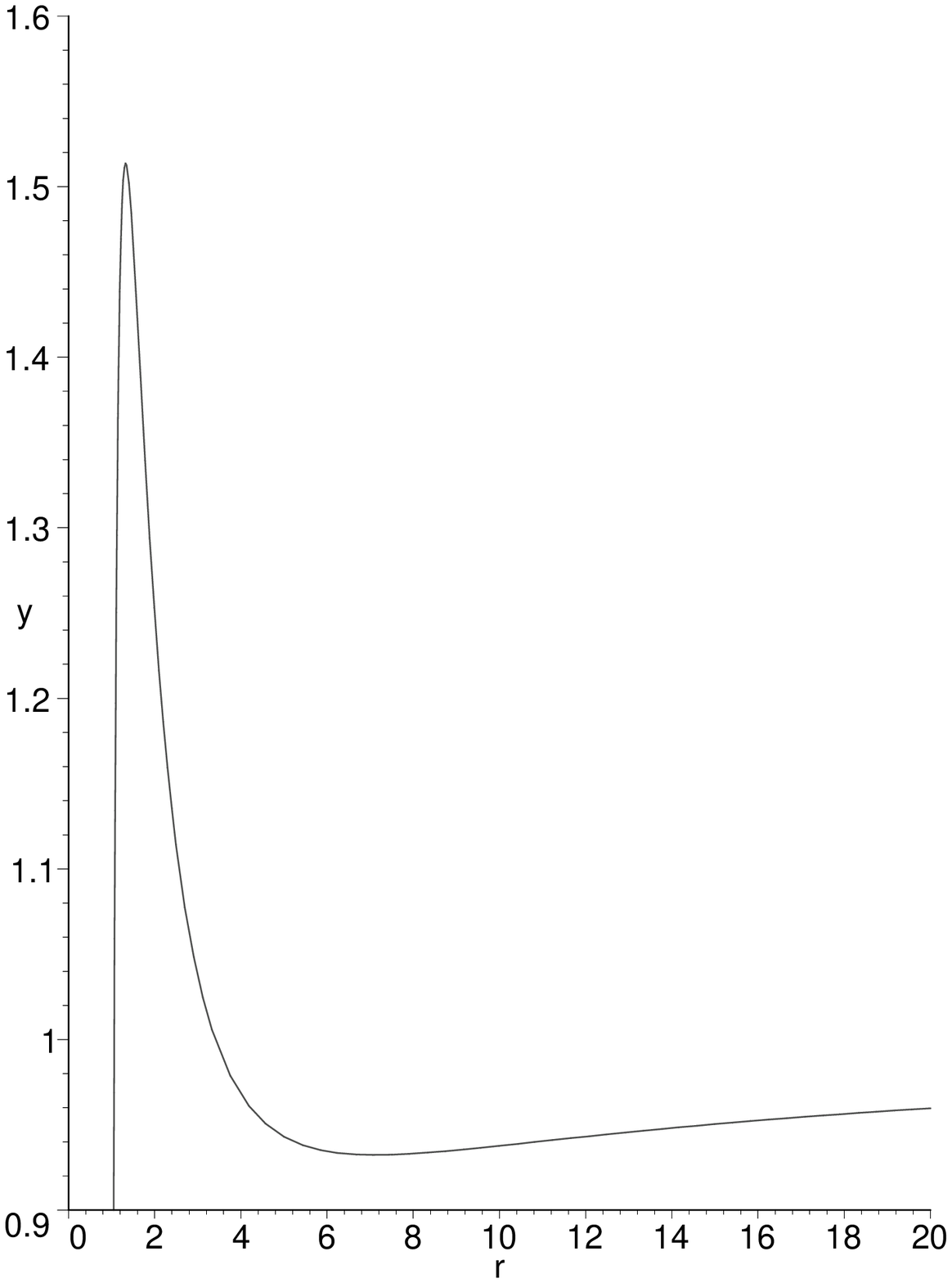} &
\psfrag{r}{$r$}
\psfrag{y}{$ $}
\epsfxsize=6.0cm
\epsfysize=6.0cm
\epsfbox{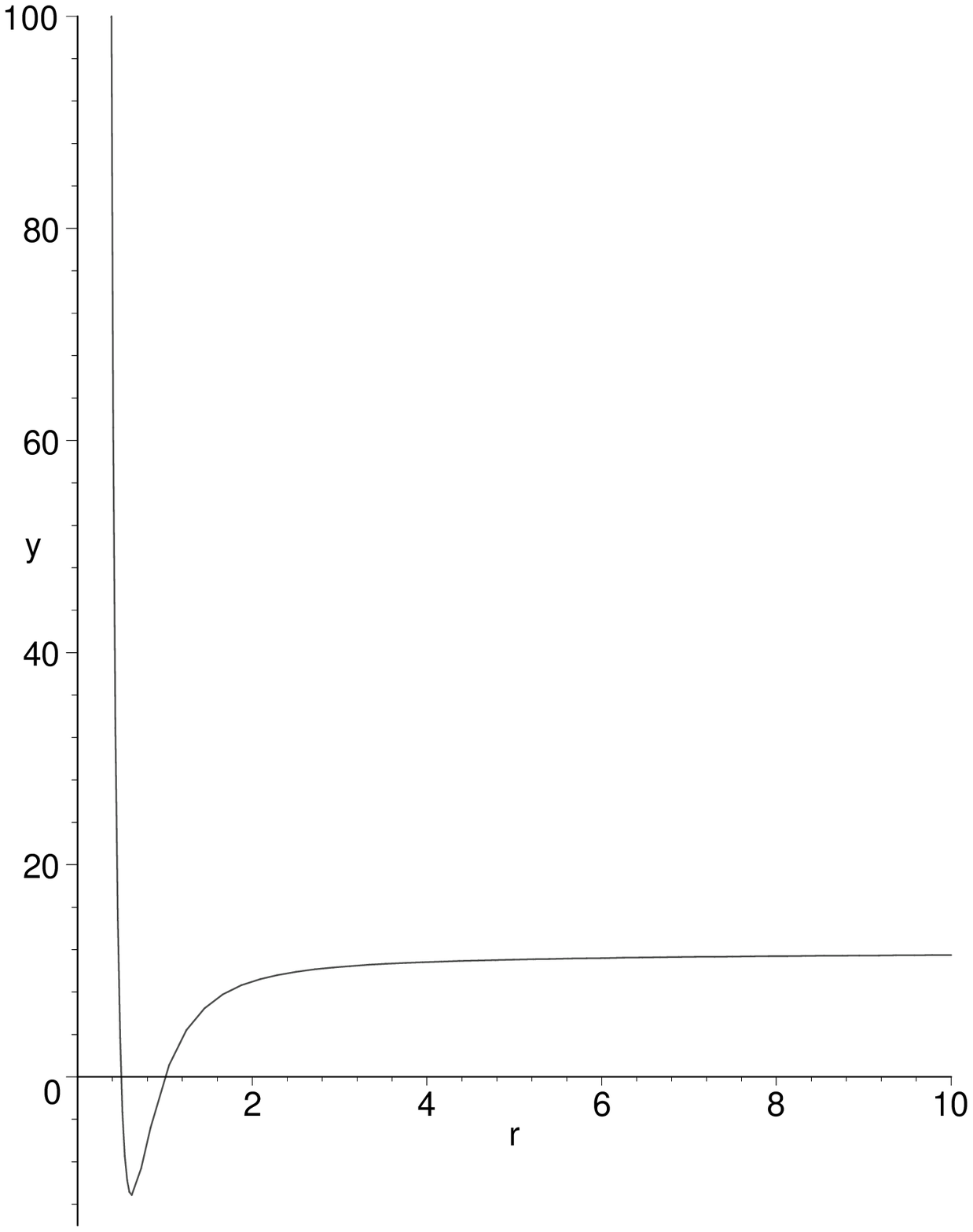}
\end{tabular}
\end{center}
\caption{In this figure we show the potentials related to axial
perturbations of $AdS$ Schwarzschild black holes. The plot on the
left represents the potential related to the evolution of an axial
spin-$\frac{1}{2}$ field, for $l=0$ and $\hat{M}=1.0$. The plot on the
right represents the potential related to the evolution of an axial
spin-$2$ field, for $l=3$ and $\hat{M}=1.0$. The masses are rescaled
with respect to the radial coordinate of the black hole event
horizons.}
\label{figu1}
\end{figure}

In Fig. \ref{figu1},
we plot the potential related to the evolution
of the axial component of massless spin-$\frac{1}{2}$ fields
alongside the axial potential for the spin-$2$ case in four
dimensional Schwarzschild-$AdS$ space-time. Note that
$r\rightarrow\infty$ is a regular point of the differential
equation \ref{even} and Dirichlet boundary conditions are a consistent choice.

\section{numerical results}
\begin{figure}[ht!]
\begin{center}
\begin{tabular}{cc}
\psfrag{wI}{$\omega_{I}$}
\psfrag{T}{$T$}
\epsfxsize=7.5cm
\epsfysize=5.0cm
\epsfbox{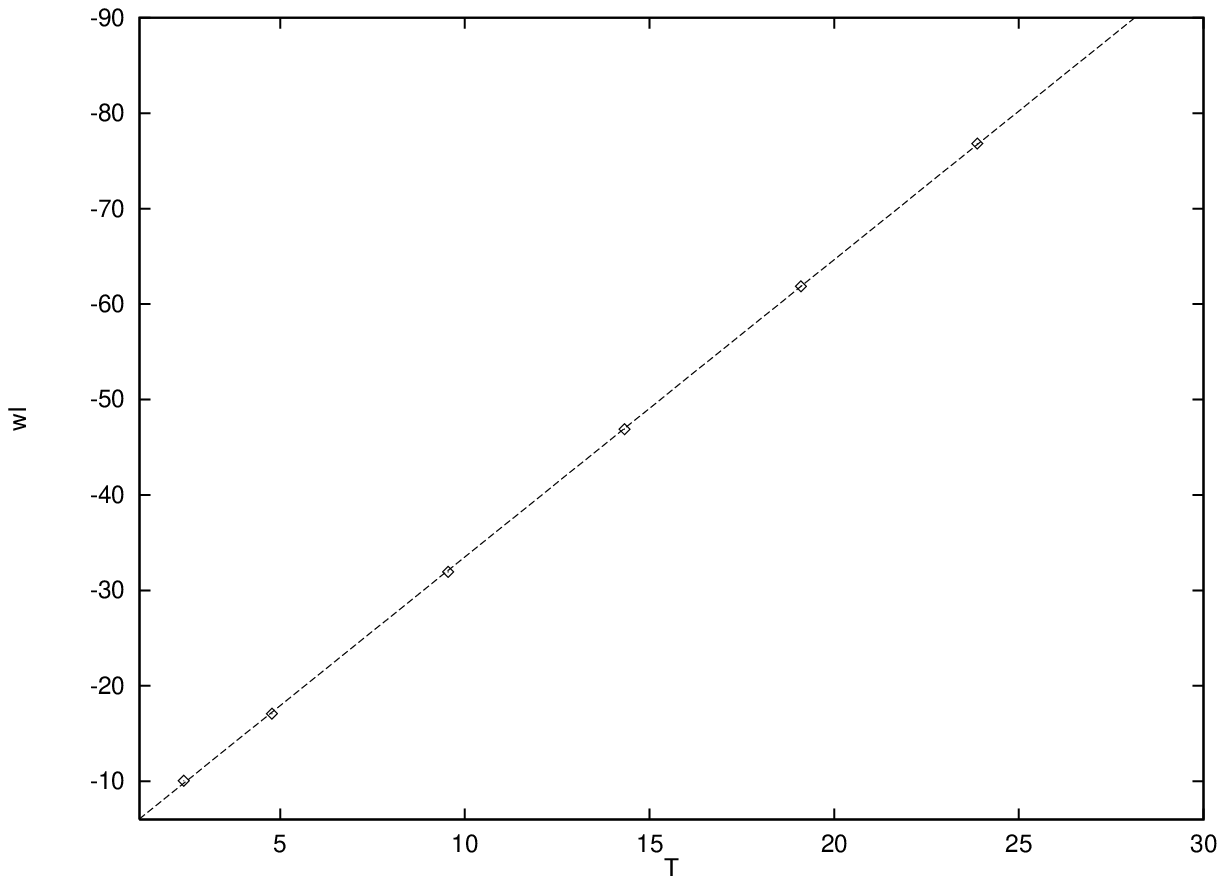} &
\psfrag{wI}{$\omega_I$}
\psfrag{T}{$T$}
\epsfxsize=7.5cm
\epsfysize=5.0cm
\epsfbox{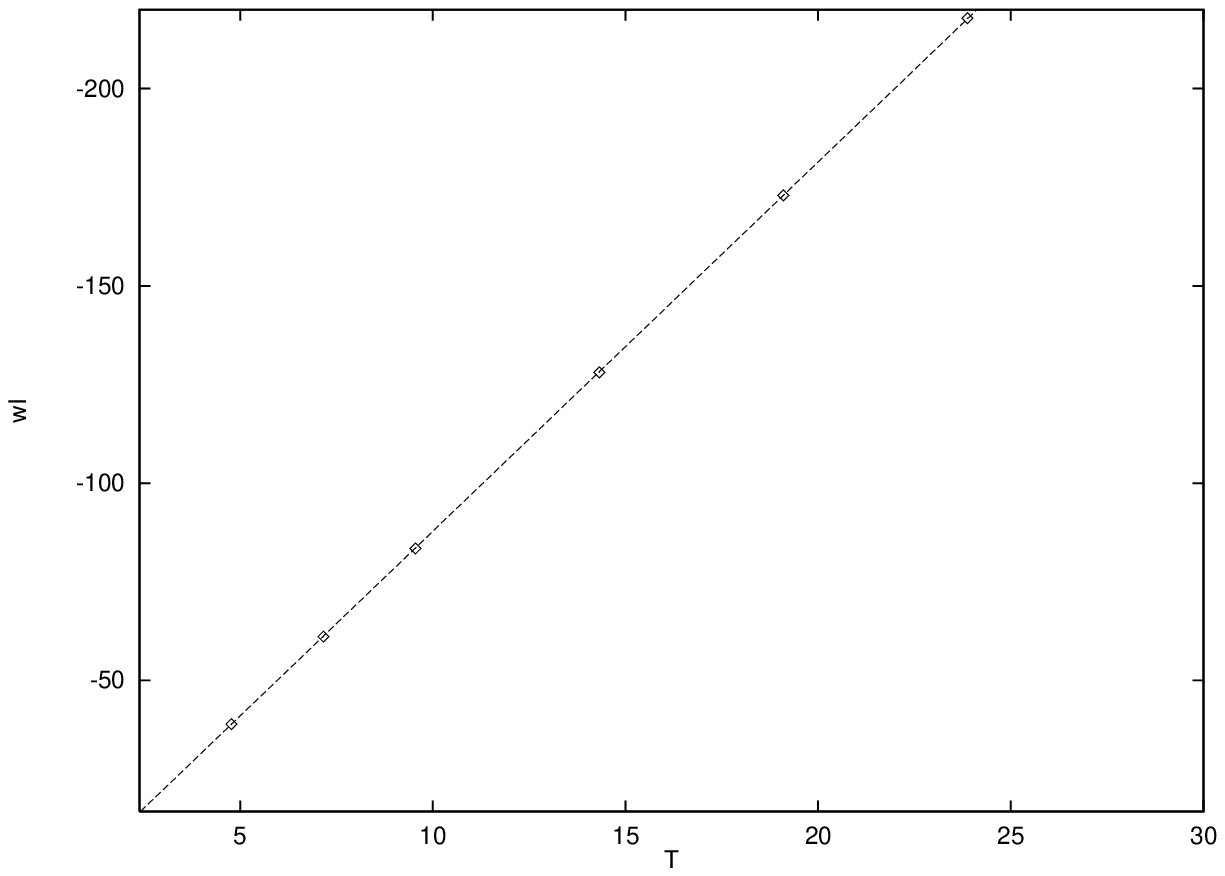}
\end{tabular}
\end{center}
\caption{The imaginary part of the spinor quasinormal mode frequencies for $n=0,  l=0$ (left) and $n=1$, $l=0$ (right) are shown as functions of the temperature $T$ for some selected black hole sizes. The dots represent some frequencies numerically evaluated. The lines connecting them are linear fits.}
\label{fig0}
\end{figure}

The numerical procedure described above can be successfully applied to both large black holes, for which the radius of the event horizon $r_1$ is much larger than the Anti-de Sitter radius $\alpha^{-1}$ $(r_{1}>>\alpha^{-1})$, and intermediate black holes for which $r_{1}\sim \alpha^{-1}$. The quasinormal frequencies are decomposed into real and imaginary parts
\begin{equation}
\omega=\omega_{R}+i\omega_{I}.
\end{equation}
With this sign chosen, $\omega_{I}$ is negative for all quasinormal frequencies. In Table \ref{table1},
\begin{table}[!ht]
\begin{tabular}{c|cc|c|cc}
\hline
\hline
$r_{1}$  &$\omega_{R}$&$\omega_{I}$&$r_{1}$&$\omega_{R}$&$\omega_{I}$\\ \hline
100&$\sim 0$&-76.8157&1.0&1.80808&-1.10565\\
60&$\sim 0$&-46.8901&0.8&1.76332&-0.8516\\
40&$\sim 0$&-31.9303&0.5&1.7463&-0.4525\\
20&$\sim 0$&-17.0652&0.45&1.7555&-0.3829\\
10&$\sim 0$&-10.0749&0.4&1.7719&-0.3118\\
5&2.20&-6.3626&0.3&1.7958&-0.1768\\
\hline
\hline
\end{tabular}
\caption{Values of the lowest Dirac quasinormal mode frequency for $\alpha^{-1}=1.0$ and $n=0, l=0$. These results are obtained by evaluating the frequency as a function of $r_{1}$ for some selected black hole sizes.}
\label{table1}
\end{table}
we list the values of the lowest quasinormal mode frequencies for $\alpha^{-1}=1$, $l=0$ and selected values of $r_{1}$. For large black holes, the real and imaginary parts of the frequency scale linearly with the horizon radius, resembling the results obtained in \cite{horowitz,lemos} for other perturbation fields. Since the temperature scales also with $r_1$ in this regime, the imaginary part of the frequency, which determines how damped the mode is and which according to the $AdS/CFT$ correspondence is a measure of the characteristic time $\tau =1/|\omega_{I}|$ of approach to thermal equilibrium, scales with the temperature. Therefore, in the dual CFT the approach to thermal equilibrium is faster for higher temperatures. The linearity of the scaling between $\omega_{I}$ and $T$ is clearly shown in Fig. \ref{fig0},
both for the $n=0$ and $n=1$ mode numbers. The dots represent some frequencies numerically calculated. The lines connecting them are linear fits. Explicitly, the lines are given by
\begin{equation}
\omega_{I}=-3.11\, T
\end{equation}
for $n=0$ and
\begin{equation}
\omega_{I}=-9.36\, T
\end{equation}
for $n=1$.

\begin{figure}[ht!]
\begin{center}
\leavevmode
\psfrag{wR}{$\omega_R$}
\psfrag{wI}{$\omega_I$}
\epsfysize=10.0cm
\epsfbox{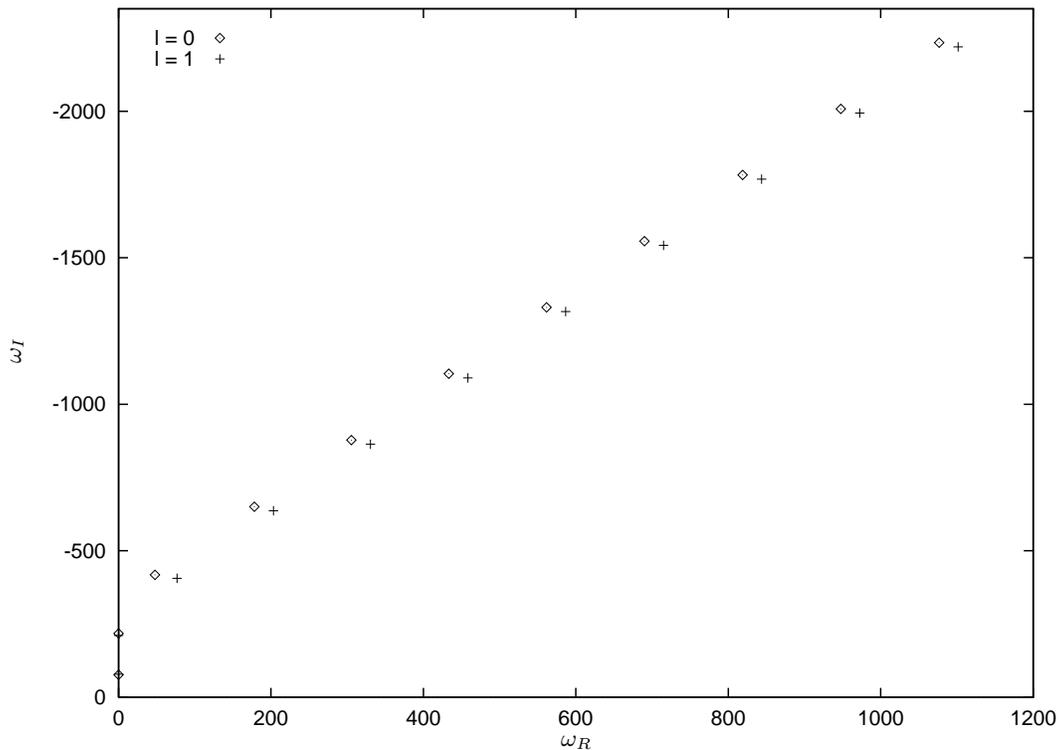}
\end{center}
\caption{The frequencies of the first eleven quasinormal modes for the
axial component of the Dirac field, are shown for $\alpha^{-1}=1$, $r_{1}=100$ and the angular mode values $l=0$ and $l=1$.}
\label{newfig}
\end{figure}

In Table \ref{table1bis},
\begin{table}[!ht]
\begin{tabular}{c|cc|cc}
\hline
\hline
Overtone Number &\multicolumn{2}{c}{$l=0$}&\multicolumn{2}{c}{$l=1$}\\ \hline
n&$\omega_{R}$&$\omega_{I}$&$\omega_{R}$&$\omega_{I}$\\ \hline
0& $\sim$ 0&-76.8402&$\sim$ 0&-78.7731\\
1&$\sim$ 0&-217.83&$\sim$ 0&-211.417\\
2&47.7354&-417.775&76.7222&-406.192\\
3&178.073&-650.494&203.247&-636.479\\
4&305.383&-877.767&330.457&-863.679\\
5&433.279&-1104.39&458.292&-1090.25\\
6&561.539&-1330.64&586.517&-1316.47\\
7&690.04&-1556.66&715.01&-1542.46\\
8&818.77&-1782.56&843.695&-1768.29\\
9&947.614&-2008.24&972.531&-1994.02\\
10&1076.56&-2233.86&1101.48&-2219.65\\
\hline
\hline
\end{tabular}
\caption{Quasinormal mode frequencies corresponding to $l=0$ and $l=1$ massless Dirac perturbations of a large Schwarzschild-AdS black hole $(r_{1}=100)$. For large values of the overtone number $n$, the modes become evenly spaced in mode number and the spacing is given by $\frac{(\omega_{n+1}-\omega_{n})}{r_1}\sim (1.299-2.25i)$.}
\label{table1bis}
\end{table}
we list the values of the first eleven quasinormal mode frequencies corresponding to massless spinor perturbations of large Schwarzschild-AdS black holes $(r_{1}\gg \alpha^{-1})$. For large values of the overtone number $n$, the frequencies become evenly spaced in mode number and the spacing, which is independent of the angular mode number $l$, is given by
\begin{equation}
\frac{(\omega_{n+1}-\omega_{n})}{r_1}\sim (1.299-2.25i).
\end{equation}
This is exactly the same spacing, related to the large black hole regime, obtained by Cardoso and his collaborators in \cite{kono}, for different kinds of perturbing fields. Their results were the same for scalar, electromagnetic and gravitational perturbations. Moreover, the quasinormal frequencies of large black holes have a number of first overtones with pure imaginary parts as in the electromagnetic and gravitational cases. The higher the black hole radius $r_1$, the higher the number of these first pure damped modes. In Fig. \ref{newfig}, the large black hole frequencies are plotted for $l=0, 1$.

\begin{figure}[ht!]
\begin{center}
\leavevmode
\psfrag{wR}{$\omega_R$}
\psfrag{wI}{$\omega_I$}
\epsfysize=10.0cm
\epsfbox{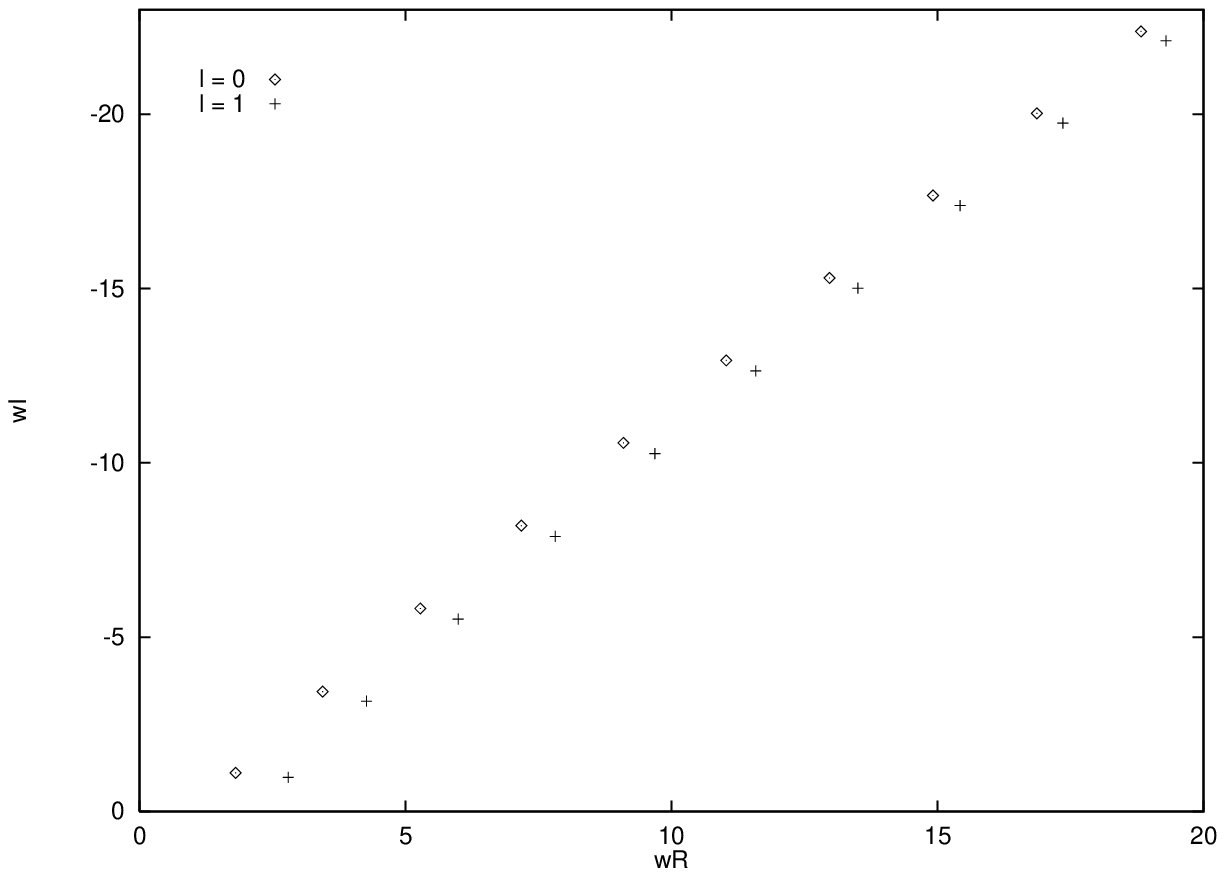}
\end{center}
\caption{The frequencies of the first ten quasinormal modes for the
axial component of the Dirac field, are shown for $r_{1}=\alpha^{-1}=1.0$, and the angular mode values $l=0$ and $l=1$.}
\label{fig2}
\end{figure}

In Table \ref{table2},
\begin{table}[!ht]
\begin{tabular}{c|cc|cc}
\hline
\hline
Overtone Number &\multicolumn{2}{c}{$l=0$}&\multicolumn{2}{c}{$l=1$}\\ \hline
n&$\omega_{R}$&$\omega_{I}$&$\omega_{R}$&$\omega_{I}$\\ \hline
0& 1.80808 &-1.10565&2.7956&-0.9744\\
1&3.44229&-3.43805&4.26976&-3.1623\\
2&5.28094&-5.82213&5.99062&-5.51618\\
3&7.17816&-8.119914&7.81556&-7.89152\\
4&9.09828&-10.5728&9.68897&-10.2668\\
5&11.0301&-12.9413&11.5837&-12.6412\\
6&12.9712&-15.3067&13.5014&-15.0117\\
7&14.9172&-17.6699&15.4250&-17.3815\\
8&16.8671&-20.0254&17.3552&-19.7468\\
9&18.8253&-22.3793&19.2938&-22.1086\\
\hline
\hline
\end{tabular}
\caption{Quasinormal mode frequencies corresponding to $l=0$ and $l=1$ massless Dirac perturbations of an intermediate Schwarzschild-AdS black hole $(r_{1}=1.0)$. For large values of the overtone number $n$, the modes become evenly spaced in mode number and the spacing is given by $\frac{(\omega_{n+1}-\omega_{n})}{r_1}\sim (1.96-2.36i)$.}
\label{table2}
\end{table}
we list the first ten quasinormal mode frequencies corresponding to massless spinor perturbations of intermediate Schwarzschild-AdS black holes $(r_{1}\sim \alpha^{-1})$. The values of the frequencies are shown for $\alpha^{-1}=1.0$ and $l=0, 1$. For large values of the overtone number $n$, the modes become evenly spaced and the spacing, which is independent of $l$, is given by
\begin{equation}
\frac{(\omega_{n+1}-\omega_{n})}{r_1}\sim (1.96-2.36i).
\end{equation}
Once again this is a result that, for the intermediate black hole regime, resembles those obtained in \cite{kono} for different kinds of perturbation fields. In Fig. \ref{fig2},
the intermediate black hole frequencies are plotted for $l=0, 1$.

\begin{figure}[ht!]
\begin{center}
\leavevmode
\psfrag{ak}{$\alpha^{-1}\kappa$}
\psfrag{awR}{$\alpha^{-1}\omega_{R}$}
\epsfysize=10.0cm
\epsfbox{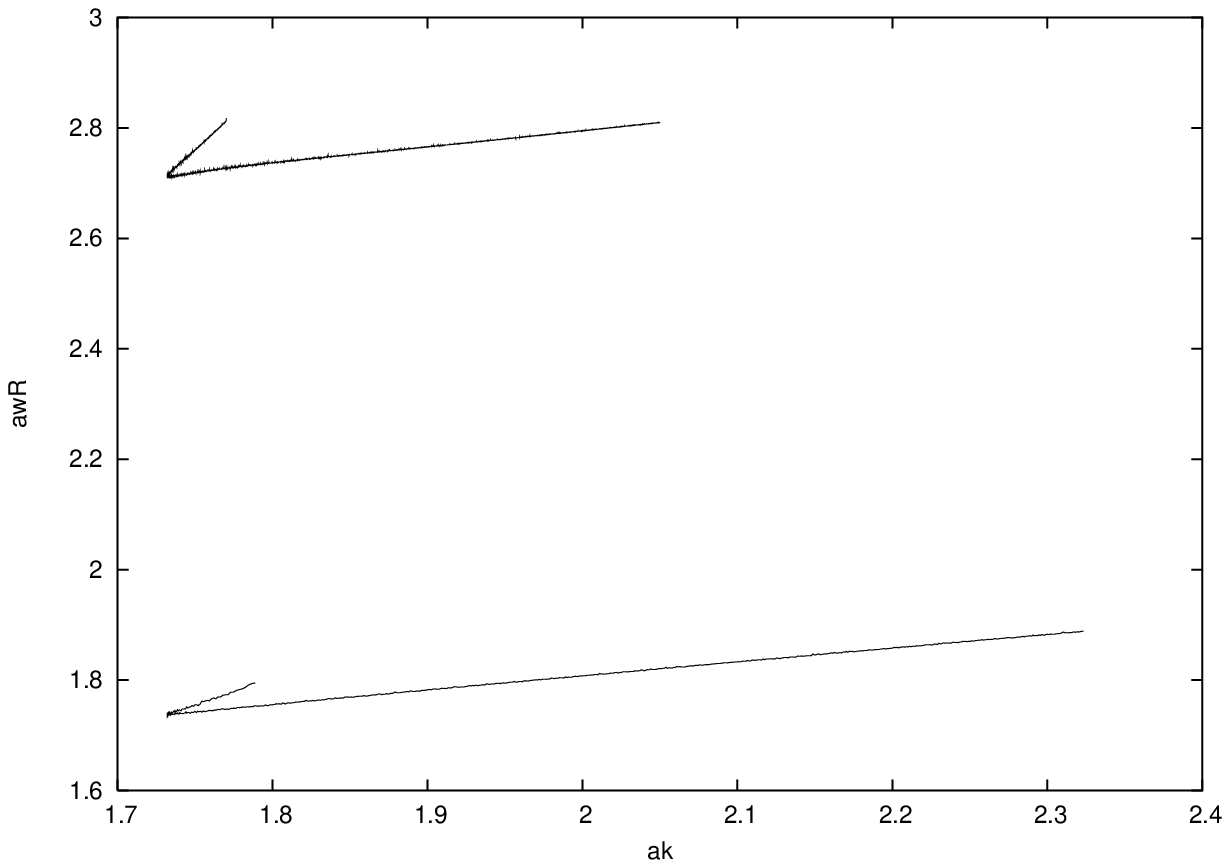}
\end{center}
\caption{The dependence of the real part of the fundamental
quasinormal mode frequencies on the metric parameters for $l=0$ (bottom)
and $l=1$ (top), is shown for Dirichlet axial boundary conditions on
the spinor field.}
\label{fig3}
\end{figure}

\begin{figure}[ht!]
\begin{center}
\leavevmode
\psfrag{alpha^-1*k}{$\alpha^{-1}\kappa$}
\psfrag{alpha^-1*wI}{$\alpha^{-1}\omega_{I}$}
\epsfysize=10.0cm
\epsfbox{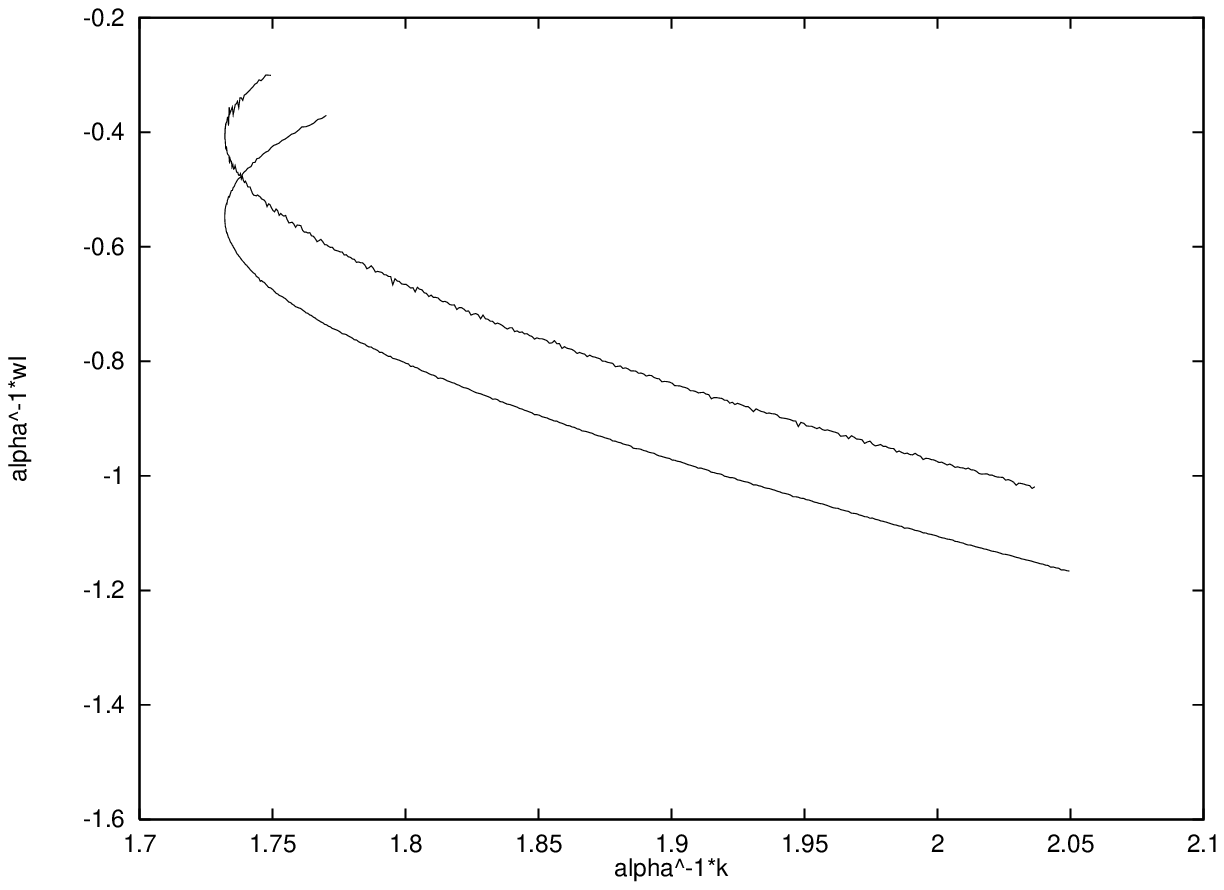}
\end{center}
\caption{The dependence of the imaginary part of the fundamental
quasinormal mode frequencies on the metric parameters for $l=0$ (bottom)
and $l=1$ (top), is shown for Dirichlet axial boundary conditions on the
spinor field.}
\label{fig4}
\end{figure}

In Fig. \ref{fig3}
and \ref{fig4},
we show, for intermediate black holes, the dependence of the real and imaginary part of the fundamental quasinormal mode frequencies on the metric parameters for $l=0$ and $l=1$. The parameters $\kappa$ and $\alpha^{-1}$ represent, respectively, the black hole surface gravity evaluated at the event horizon and the Anti-de Sitter radius. Both the real and imaginary parts of the frequencies seem to depend linearly on the black hole surface gravity and therefore on the black hole temperature for $\alpha^{-1}\kappa>1.8$. This is a result that has already been found in \cite{moss} for gravitational perturbations of Schwarzschild $AdS$ black holes. The slopes of the straight lines, which are independent of $l$, are given by
\begin{equation}
\omega_{R} = 0.25 \, \kappa
\end{equation}
for the plots in Fig. \ref{fig3} and
\begin{equation}
\omega_{I} = -1.29 \, \kappa
\end{equation}
for the plots in Fig. \ref{fig4}.

For small black holes $(r_{1}\ll\alpha^{-1})$, it is very difficult to evaluate the corresponding quasinormal frequencies, or to go high in mode number. The error associated in estimating the frequencies in this regime is too high, and we cannot be completely sure of the results. In the case of other perturbation fields
\cite{horowitz, lemos, kono}, it was found that small black holes have quasinormal frequencies that are very close to the pure $AdS$ values. In the case of the Dirac field considered in this paper, after setting $\alpha^{-1}=1$, we have been able to evaluate the fundamental frequency, at most, down to the value of $r_{1}=0.3$ (see Table \ref{table1}). Further investigations and more powerful numerical tools are needed to outline the behaviour of the spin-$1/2$ frequencies in the limit of $r_{1}\rightarrow 0$.

\section{conclusion}
\label{section4}
In this paper we have investigated the quasinormal modes frequencies
of a spinor field interacting with a Schwarzschild black hole in four
dimensional Anti-de Sitter space-time. After solving the related
axial perturbative equation we have imposed on this solution
Dirichlet boundary conditions and evaluated the corresponding
quasinormal frequencies. This choice of boundary conditions is
physically justified in the contest of the $AdS/CFT$ correspondence,
in which it is suggested to exploit these frequencies in the
evaluation of poles of correlation functions associated to the
conformal field theory on the boundary.

For large black holes, we have numerically confirmed that the imaginary part of the frequency scales with the temperature. We have also shown that both the real and imaginary components of the frequencies are evenly spaced in mode number. The spacing between consecutive modes, which is independent of the angular quantum number $l$, behaves as in the case of scalar, electromagnetic and gravitational perturbations. The quasinormal frequencies have a number of first overtones with pure imaginary parts.

For intermediate black holes, we have found that both the real and imaginary part of the frequency scale linearly with the surface gravity. The modes are evenly spaced and the spacing, which is independent of the angular quantum number $l$, resembles the one related to other kinds of perturbations.

Our purpose is to extend this project imposing different boundary
conditions on the axial or the polar Dirac equations. We could be
interested, for instance, in calculating the correlation functions
and associated poles of a quantity related to the axial part of the
spinor field. That would suggest we set the polar component of the
spinor field perturbation to vanish on the boundary and consequently
we could solve either the odd parity differential equation with
Dirichlet boundary conditions or the even parity equation with
mixed boundary conditions due to the transformation theory
between polar and axial functions.
\begin{acknowledgments}
M. Giammatteo would like to thank I. Moss for helpful suggestions
and R. Malagigi for useful discussions. J. L. Jing was supported
by the National Natural Science Foundation of China under Grant
No. 10275024; the FANEDD under Grant No. 2003017.
\end{acknowledgments}

\end{document}